\documentstyle[prl,aps,epsf,twocolumn]{revtex}

\tighten

\begin{document}
\draft

\draft
\twocolumn[\hsize\textwidth\columnwidth\hsize\csname@twocolumnfalse\endcsname

\title{\bf 
 Activated
mechanisms in amorphous silicon: an activation-relaxation-technique
study}

\author{Normand Mousseau\protect\cite{mousadd}}

\address{Department of Physics and Astronomy and Condensed Matter and Surface
Science Program, Ohio University, Athens, OH 45701, USA}

\author{G.T. Barkema\protect\cite{barkadd}}

\address{Theoretical physics, Utrecht University, Princetonplein 5,
3584 CC Utrecht, the Netherlands}

\date{\today}

\maketitle

\begin{abstract}
At low temperatures, dynamics in amorphous silicon occurs through a
sequence of discrete activated events that locally reorganize the
topological network. Using the activation-relaxation technique, a data
base containing over 8000 such events is generated, and the events are
analyzed with respect to their energy barrier and asymmetry,
displacement and volume expansion/contraction.  Special attention is
paid to those events corresponding to diffusing coordination defects.
The energetics is not clearly correlated with the displacement, nor
with the defect density in well relaxed configurations. We find however
some correlation with the local volume expansion: it tends to increase
by about 4 eV/\AA$^3$.  The topological properties of these events are
also studied; they show an unexpectedly rich diversity.
\end{abstract}


]                                                           

\vspace*{-0.5cm}


\section{Introduction}

The properties of amorphous semiconductors can vary widely as a function
of the details of the preparation method: hot-wire, electron-beam
deposition\cite{sputt} and ion bombardment\cite{roorda91} can yield samples
with a significant spread in electronic and structural properties.  Such
diversity is a reflection of the immensely large number of metastable
configurations of nearby energy.

The topological structure of this complex energy surface can be sampled
indirectly, for instance by light illumination or ion bombardment,
bringing a sample from one metastable state to another, often in a
(statistically) reversible manner.  A direct study of this energy
surface requires the identification at the microscopic level of the
mechanisms responsible for moving from one metastable state to another,
and is much harder to perform.  Because of the high degree of disorder,
very few techniques can provide a truly microscopic representation of
the bulk dynamics~\cite{mousseau97}. At best, one can extract some
quantity averaged in time and space, providing a very rough picture of
what is really happening.

In spite of these difficulties, the past few years have seen
significant experimental and theoretical efforts to try to provide some
insight into the bulk dynamics.  As it is becoming evident that little
hard and precise information about the local environment can be
obtained by via static methods, more and more emphasis is put onto the
development of techniques to sample dynamical quantities.

This article presents the first detailed study of the microscopic
nature of activated mechanisms in {\it a}-Si. The method that we have
used is the activation-relaxation technique (ART), introduced by us a
few years ago \cite{barkema96}. Here, we apply it to an empirical model
of amorphous silicon as described by a modified Stillinger-Weber
potential \cite{weber85}. The activation-relaxation technique allows one to
concentrate on the activated mechanisms that are responsible for most
of the dynamics below melting. We have already reported a first stage
of this study in a recent Letter \cite{barkema98}, where we
concentrated on only those mechanisms involving no coordination
defects.  Here we look at a wider spectrum of mechanisms, with a special
emphasis on the diffusion of coordination defects.

This paper is organized as follows.  In section \ref{method} we describe
the activation-relaxation technique and give the details of the
simulations that result in the data base of events.  Next, in
section~\ref{results}, we present the results extracted from this data base.
Because this type of work is rather new, we also discuss the analysis
of the data in some detail.  The results contain both a global analysis
and a more detailed topological classification.

\section{Method of generation of the event database}
\label{method}

In this article, we concentrate on identifying and classifying activated
mechanisms that are responsible for the relaxation and diffusion in
amorphous silicon.  This is done using the activation-relaxation
technique (ART), an energy-landscape method, that searches for barriers
and new states in complex landscapes. 

As we shall see, the results we obtain are in general agreement with
many of the experimental results mentioned above and provide some bounds
on the type of mechanisms that can take place in {\it a}-Si.  Of course,
empirical potentials have strong limitations, especially far away from
the equilibrium position for which they are developed. Without giving
too much weight to the exact numerical values of the activation
energies, it is nevertheless possible to give a first broad picture of
the wide variety of mechanisms that can be associated with diffusion and
structural relaxation. To go beyond the results presented below, it will
be generally necessary to use more accurate interactions such as
tight-binding or plane wave methods.

\subsection{sampling one event with ART}

The aim of ART is to sample minimum-energy paths, starting in a local
energy minimum, passing through a first-order saddle point (where the
minimum-energy path has its highest point), and leading to another local
energy minimum. ART does this in three stages: leaving the so-called
``harmonic well'', convergence to the saddle point, and convergence to
the new minimum.  The last stage, relaxation to a local energy minimum,
is straightforward and can be achieved by a wide range of standard
minimization techniques, for instance the conjugate gradient (CG) method
\cite{numrec}. The first two stages represent the activation to a
saddle point and are specific to ART.  As the end conditions of the first
stage (leaving the harmonic well) are set by the actual implementation
of the second stage (finding the saddle point), we will discuss these two
stages in reversed order.

The second stage is convergence to a first-order saddle point. At such a
saddle point, the gradient of the energy is by definition zero in all
directions, and the second derivative of the energy is positive in all
directions but one. The single direction with negative curvature is
that along which the minimum-energy path proceeds. Within ART, we make
the assumption that the direction of the minimum-energy path in the
saddle point and the direction towards the original local energy
minimum have a significant overlap, i.e. that their dot-product is
significantly non-zero. If this assumption holds, a modified force
field can be introduced in which the saddle point is a
minimum~\cite{assumption}. This
field is defined by the force ${\bf G}$:
\begin{equation}
{\bf G} = {\bf F} - \left[1 - \alpha/ (1+\Delta x) \right] \left( {\bf F}
\cdot {\bf \hat{\Delta x}} \right) {\bf \hat{\Delta x}},
\end{equation}
where ${\bf F}$ is a $3N$-dimensional force vector obtained from the
first derivative of the potential energy, ${\bf \Delta x}$ is the
displacement vector from the minimum, and $\alpha=0.15$ a parameter determining how fast
the motion is to the saddle point. In this second stage, the redefined
force ${\bf G}$ is followed iteratively, usually along conjugate
directions, starting from just outside the harmonic well around the
original local energy minimum.  Ideally, this process would bring
configuration directly to the saddle-point and stop there, but because
the projection is an approximation of the valley, the configuration
passes in the viccinity of saddle point without halting. We therefore
stop the activation as soon as the component of the force ${\bf F}$
projected onto the displacement ${\bf \Delta x}$ changes sign, an
indication that a saddle point has just been past.  At that point, the
activation is stopped, the configuration stored as the activated
configuration and we move to the relaxation. A more accurate convergence
to the saddle point can be obtained by following directions determined
by the eigenvectors of the dynamical Hessian \cite{doye97}, but this is
too costly for the system size of interest here.

Most saddle points, even those belonging to energetically favorable
minimum-energy paths, cannot be reached by following this redefined
force ${\bf G}$ from a point well inside the harmonic region, i.e., the
region of the energy landscape that is well approximated by a
high-dimensional parabola centered on a local minimum.  We refer the
reader to Ref.~\cite{mousseau98} for a detailed discussion of the
origin of this problem and just describe our algorithm. We thus have to
make sure that the configuration has left the harmonic well before
following ${\bf G}$.  This first stage is implemented as follows.  In a
local energy minimum configuration, a few atoms and their nearby
neighbors are displaced randomly. The total displacement is small,
typically 0.01 \AA, and serves to create a non-zero force. At this
point, the force will mostly point back to the minimum, and the
component of the force, parallel to the displacement, will dominate the
perpendicular components. However, we then increase the displacement
from the original minimum until this no longer holds, and consequently
we are outside the harmonic region. At that point, we can start the
second stage of the activation and follow ${\bf G}$.

\subsection{generating the data base}

The database of events that we use in the current manuscript is
generated as follows. The energy landscape is described by the
Stillinger-Weber potential \cite{weber85}, modified as described below.
This empirical interaction includes a two-body and a three-body
interaction:
\begin{equation}
E =  \sum_{\langle ij \rangle} V(r_{ij}) + \sum_{\langle ijk \rangle}
V(r_{ij},r_{ik},\theta_{jik})
\end{equation}
where the brackets $\langle$ and $\rangle$ indicate that each bond or angle
is only counted once; the two-body potential is
\begin{equation}
V(r_{ij}) = \epsilon A \left(Br_{ij}^{-p}-1\right) \; 
\exp\left[(r_{ij}-a)^{-1}\right]
\end{equation}
and the three-body potential is
\begin{eqnarray}
V(r_{ij},r_{ik},\theta_{jik})& =
\epsilon \lambda \;
\left( \cos \theta_{jik} + \frac{1}{3} \right)^2 \nonumber\\
\times & \exp \left[\gamma (r_{ij}-a)^{-1}\right]
         \exp \left[\gamma (r_{ik}-a)^{-1}\right]
\end{eqnarray}
 
The numerical values for the parameters are $A=7.050$, $B=0.6022$,
$p=4$, $a=1.80$, $\lambda=31.5$, $\gamma=1.20$, $\sigma=2.0951$\AA\,
and $\epsilon=2.1682$ eV;
this set is identical to that used by Stillinger and Weber except for
$\lambda$ which has been increased by a factor of 1.5 in order to
provide a more appropriate structure for {\it a }-Si
\cite{barkema96,mousseau98,ding86}. Recent work on fracture
underlines the fact that none of the available empirical interaction
potentials describe silicon exactly~\cite{hauch99}. This is particularly
the case for energy barriers and density of defects; trends, more than
exact values, are therefore what we are looking for here. 

We report here on results obtained from three independent runs. Each
initial 1000-atom cell is constructed by a random packing in a large
cubic cell. This configuration is then minimized at zero pressure to a
nearby minimum state.  ART is then applied iteratively with a
Metropolis temperature of 0.25 eV in order to bring the configuration
to a well-relaxed amorphous state; we consider that a system is
``well-relaxed'' when the energy does not decrease significantly over
hundreds of events.  This takes place after about 5000 trial events,
with a success rate slightly above 65\%.

\begin{figure} 
\epsfxsize=8cm
\epsfbox{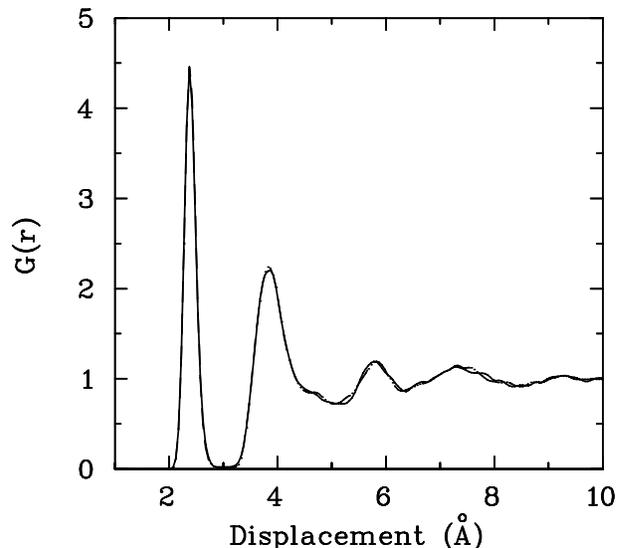}
\caption{ Smoothed radial distribution function for configurations C1
(solid line) and C5000 (dashed line)} 
\label{fig:rdf} 
\end{figure}

After reaching a plateau in energy, for each event we store the initial
minimum configuration, the saddle point configuration, and the final
minimum configuration. Over the three independent runs, we collected a
set of 8106 events.  Figure \ref{fig:rdf} shows the radial distribution
function for run C at the beginning (C1) and end of the data
acquisition run, 5000 trial-events later (C5000). Although radial
distribution functions cannot discriminate easily between realistic and
non-realistic structure \cite{mousseau97b}, the one obtained here is in
good agreement with experimental data \cite{rdf-exp}.

Table \ref{tab:rdf} shows the structural properties of these two
networks.  A third of the bonds, involving 42\% of the atoms, have been
changed and yet the total energy and structural properties are almost
unchanged. This is an indication that the configuration, through a
sequence of events, has evolved considerably on an almost constant
energy surface.

\begin{table}

\caption{Structural properties of configurations C1 and C5000, the first
and last structures used during the data collection in run C:
the energy per atom in electron-Volt, the distribution of coordination using
a cut-off at 3.05\AA, the average bond angle and the width of the
related distribution.  The two configurations differ by 33\% in their
bonding and 42\% of the atoms have changed their neighbor list. }
\begin{tabular}{rrr}
                    & C1     & C5000     \\ \hline 
Energy (eV)         & -3.903 & -3.918 \\  
3-folds             &  0.043 &  0.035 \\
4-folds             &  0.948 &  0.960 \\
5-folds             &  0.009 &  0.005 \\
$\langle r \rangle$ &  3.966 &  3.97  \\
$\theta$            & 109.27 & 109.32 \\
$\Delta \theta$     &  10.04 &   9.74 \\
\end{tabular}
\label{tab:rdf}
\end{table}

The bond angle distribution and coordination is comparable with the best
networks as obtained with realistic potentials \cite{ding86,bazant}.
A comparison with experiment is more difficult.  Experimental works on
amorphous silicon samples prepared by ion-bombardment report that
homogeneous samples of {\it a}-Si, without voids, have a density about
1.8 \% lower than {\it c}-Si \cite{custer94,williamson95}.  Recent
high-Q X-ray diffraction measurements on similarly prepared samples show
that well-relaxed {\it a}-Si could have an average coordination as low
as 3.88, much below the 4.0 generally considered to be appropriate for
ideal {\it a}-Si~\cite{laaziri}. The authors of this work conclude that
the high density of dangling bonds should be responsible for the lower
density of these samples. This analysis is not supported by our
simulation, which leads to cells with a density as low as 7 \% below that of
the crystal, while keeping the number of defects {\it lower} than that
seen in this experiment.  Clearly more work needs to be done to clarify
this situation, especially since the 12 \% of dangling bonds seen by
high-Q diffraction is at least an order of magnitude higher
than what can be expected from either differential scanning calorimetry
(1\% defects) and electron-spin resonance measurements (0.04\% defects)
\cite{roorda91}.  These results also contrast with ab-initio and
tight-binding computer simulations, which tend to find higher
coordination, often above 4.0 \cite{simulations}. The origin of this
discrepancy is hard to identify at the moment but it can be due to a
combination of inaccurate interactions and/or the differences in time
and lengths scales between simulations and experiments. In the case of
our simulations, the stability of the configurations after 5000
trial-events suggests that we have reached some type of thermalization
for this given interaction potential. 

\section{General properties of events}
\label{results}

\subsection{asymmetry, activation energy, and displacement}
\label{asym}

\begin{figure}
\begin{center}
\epsfxsize=8cm
\epsfbox{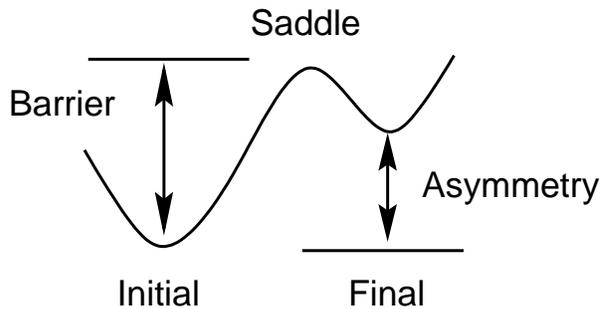}
\end{center}
\caption{The energy barrier and asymmetry in a schematic event as created
by the activation-relaxation technique.
\label{fig:barrier}}
\end{figure}

The events can first be classified in term of three energies: the energy
asymmetry, the activation energy, and the total atomic displacement (see
Fig. \ref{fig:barrier}.) The simplest quantities to use for the
classification of events are the barrier and asymmetry energies. In
Ref.~\cite{barkema98}, we give the distribution of these two quantities for
the full set of 8106 events. Both distributions are wide and relatively
smooth. To push further the analysis, it is useful to establish a first
classification based on topological properties of the network. 

The radial distribution function of relaxed {\it a}-Si goes to zero
between the first and second neighbor. The middle of this region
between first and second-neighbor lies at around 3.05 \AA.  This allows
us to establish a clear definition of nearest-neighbor bonds between
atoms.  Structural properties of the first and last minimum-energy
configurations of run C in our database are listed in Table
\ref{tab:rdf}; we find similar numbers for runs A and B. Most atoms that
change neighbors in an event are four-fold coordinated both before and
after. We discern three topological classes of events:  if {\it all}
atoms involved in an event keep their coordination unchanged between the
initial and final state, we term it a {\it perfect} event; if
topological defects change place during an event, but their total number
is conserved, we have a {\it conserved event}, which describes defect
diffusion; the remaining events are called {creation/annihilation}
events.

The exact details of the number of each class of events depends
slightly on the cut-off radius between first and second neighbors;
although low-energy configurations have a clear gap between first and
second neighbor peaks in the radial distribution function, many saddle
and some high-energy final configurations show structure in this gap.
For consistency,  we have chosen a fixed cut-off at 3.05 \AA\, for all
our analysis; the topological classification of some events might be
affected by these parameters but the overall conclusions are not
sensitive to the fine tweaking of that value.      

In a previous letter, we have discussed in some detail the class of
perfect events (802 events) \cite{barkema98}.  The bulk of our database
comprises creation/annihilation events, with 5325 events, but until now
we did not succeed in revealing interesting characteristics from
these.  In this paper, we focus on the 1979 conserved events in
our database, describing directly the diffusion of defects without
the creation or annihilation of coordination defects.

\begin{figure}
\epsfxsize=8cm
\epsfbox{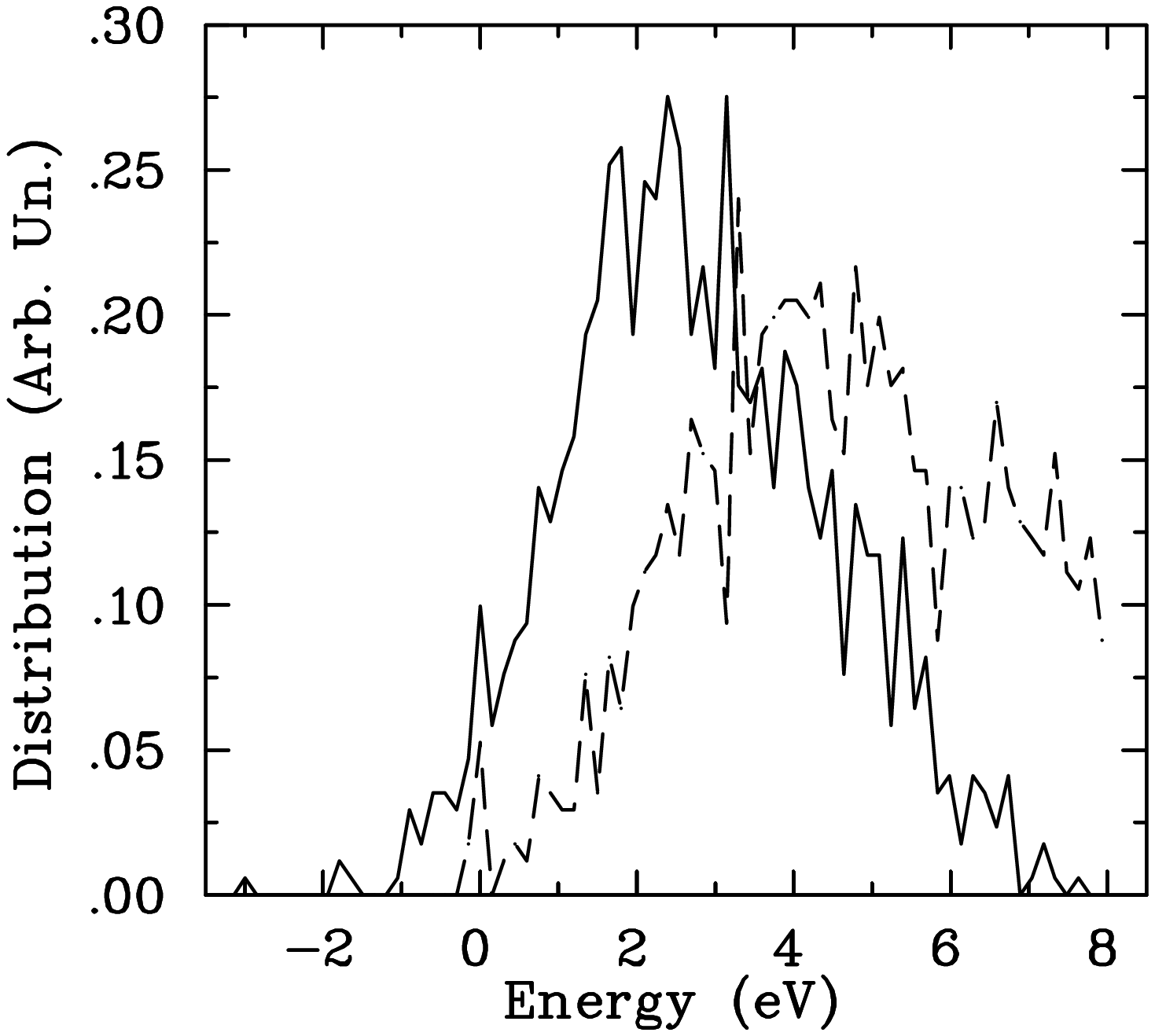}
\epsfxsize=8cm
\epsfbox{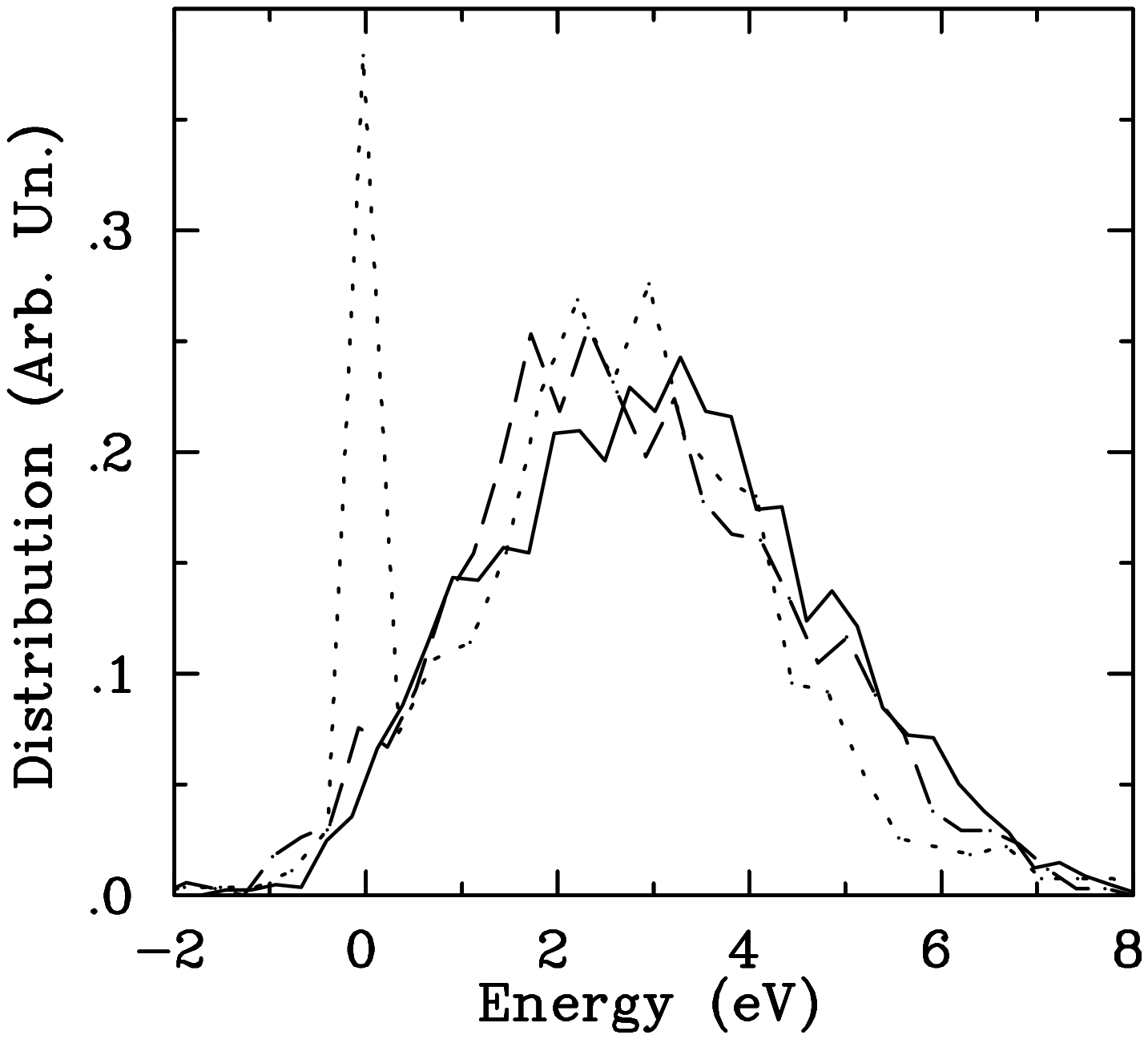}
\caption{
Top: Distribution of the energy barrier (dashed line) and energy
assymetry (solid line) for the 1979 conserved events. Bottom: Comparison
of the renormalized asymmetry energy distribution for 
the conserved (dashed line) and the perfect events
(dotted line) and all the others (solid line). }
\label{fig:energy}
\end{figure}

As already mentioned, we produced 1979 conserved events, i.e., events
where the number of coordination defects is identical in the initial
and final state. The distribution of barriers and minimum to minimum
energy differences is shown in Figure \ref{fig:energy}(a). The front
of the barrier distribution peaks at about 4.5 eV while the asymmetry
peaks at about 2.1 eV.  This distribution is very similar to the total
distribution shown in Fig.~1 of Ref.~\cite{barkema98}.  In fact, one of
the most striking results we obtain is that the distribution of barriers
and asymmetry is almost independent of the subset of events we select.
The bottom box of Fig. \ref{fig:energy}, for example, compares the
asymmetry distribution for the three classes of events: perfect,
conserved, and creation/annihilation.  Except for the peak at 0 eV in
the case of perfect events, involving atomic exchanges without
modification of the overall topology of the network, the three
distributions fall almost on top of each other at low asymmetry.  The
maximum of each distribution is slightly shifted, however, with peaks
at about 2.3 eV for the conserved events, 2.5 eV for the perfect events
and 3.2 eV for creation/annihilation events.

The bias towards higher energies in the asymmetry distribution is
expected: the distribution includes all attempted events, and not just
those that are accepted; since the run is started in an already
well-relaxed configuration, most events lead to higher energy
configurations. 

A similar insensitivity can be seen with the activation barriers. 
Although the precision on the barrier height is about 0.5 eV within the
modified-force approximation, we can still say something about
activation.  Fig. \ref{fig:energy} also plots the distribution of
barriers. It peaks at about 4.0 eV Taking into account the uncertainty
on the empirical potential, this result are consistent with experimental
measurement.  Shin and Atwater \cite{shin93} conclude, based on
conductivity measurements, that the activation-energy spectrum extends
from as low as 0.25 eV to about 2.8 eV, supposing that a prefactor
(entering logarithmically in the relation) is of order 1.  Using
isothermal calorimetry, Roorda {\it et al} \cite{roorda91} find
relaxation with a characteristic time of about 110 s between 200 and 500
C, also indicating a high activation barrier. Moreover, these results
seem to depend only weakly on the method of preparation (vacuum
evaporation or ion implantation) and are limited by the fact that above
500 C the samples tend to crystallize.  Both results are therefore also
consistent with a continuous distribution of activation barriers.     

The tail of the distribution goes much beyond experimental values, and
extends past 20 eV. Although such mechanisms are clearly unphysical,
they underscore the fact that ART does not suffer from slowing down as
the height of the barrier increases. This method is perfectly at ease
with barriers of 0.1 eV as well as those of 25 eV. In the rest of this
paper, we will concentrate on events with the 1147 more physical
barriers of less than 8 eV.  

\begin{figure}
\epsfxsize=8cm
\epsfbox{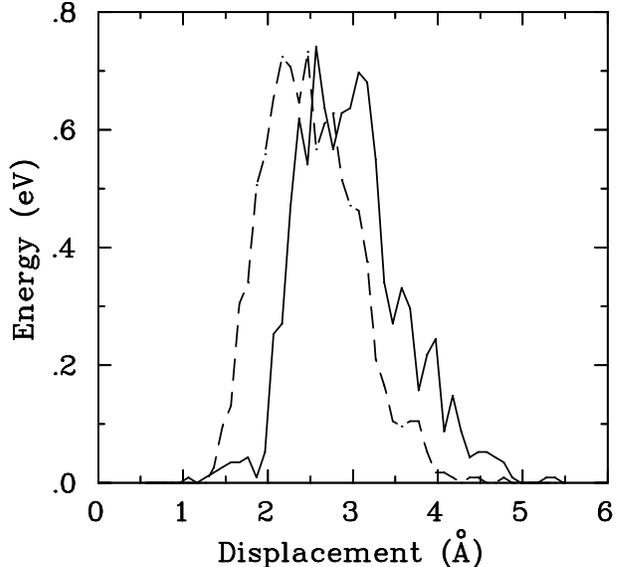}
\caption{Distribution of total displacement at the barrier (dashed line)
at at the new minimum (solid line) for conserved events. }
\label{displacement}
\end{figure}

Figure \ref{displacement} shows the histogram of the total
displacement, defined as the square root of the sum of the square of
each single-atom displacement, for these conserved events with a barrier
of less than 8 eV.  There is again little structure in the distribution.
We note that the the average displacement to the saddle point is shorter
than that to the new minimum. Based on preliminary simulations in other
materials, this trend, although intuitively reasonable, is not always
present and might be indicative is certain type of activation; more work
remains to be done to clarify this issue.

\subsection{Volume expansion/contraction per event}
\label{bvol}

If is often suggested that there should be correlation between the
energy of an event and its size. This immediately raises the question as
to whether mechanisms by which the structure rearranges itself are local
or non-local.

Based on the observation that isothermal heat release curves obey
bimolecular reaction kinetics and that the ion-beam-induced derelaxation
scales with the density of displaced atoms due to the ion bombardment
and appears to be independent of electronic energy-loss mechanisms,
Roorda \cite{roorda91} concludes that point-defect annihilation should
control the structural relaxation. It remains unclear, however, whether
this means the actual removal of defects or simply a clustering or a
passivation by hydrogen atoms.  A similar relaxation (to within a factor
2) of ion-bombarded {\it c}--Si and {\it a}--Si suggests that both
materials have similar relaxation mechanisms but not necessarily
identical. This similarity would point towards relatively local
mechanisms of defect diffusion and relaxation since at this length
scale, crystalline and amorphous materials resemble each other closely.

Based on EPS data however, Muller {\it et al.} suggest that from 1000
to 10 000 atoms have to move marginally in order to enable a single
dangling-bond defect to anneal \cite{muller94}.  This would be, at
least qualitatively, in agreement with XPS measurements that suggests
that the formation of dangling bonds in {\it a}-Si:H under exposure to
light is also accompanied by long-range structural rearrangements of
the amorphous network~\cite{masson95}, but it is in clear disagreement with
the conclusion of Roorda {\it et al.} mentioned above.  The annealing
mechanism suggested by this group, for example, is mutual annihilation
of low- and high-density defects - vacancy/interstitial.  Roorda {\it
et al.} propose defects similar to that of {\it c}-Si but not
necessarily identical~\cite{roorda91}.  M{\"o}ssbauer experiments with
Sn suggest that vacancies exist also in a-Si~\cite{liang94}. Part of
the difficulty in assessing more clearly the type of defects involved
in relaxation and diffusion is that amorphous silicon crystallizes at
about 500 C, rendering studies of self-diffusion very
difficult~\cite{polman90}.             

Because it is not always clear what size means, we consider here three
definitions: number of atoms, total displacement and local
density deformations. 

\begin{figure}
\epsfxsize=8.0cm
\epsfbox{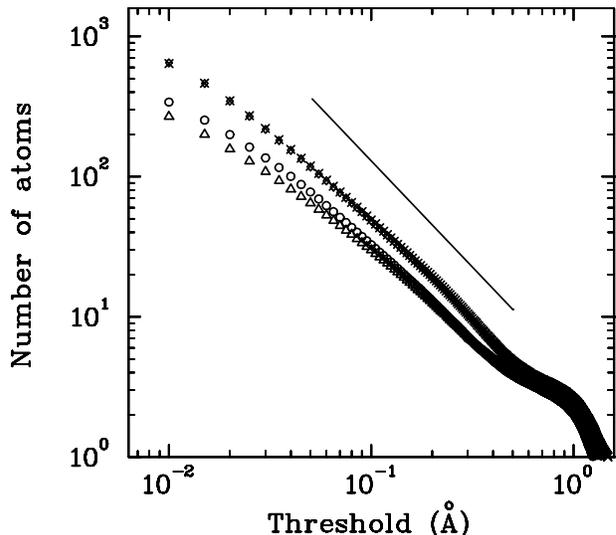}
\caption{Number of atoms displaced by a minimum threshold distance.  The
triangles and the X's are for saddle and new minimum position averaged
over all events; the circles and diamonds are averaged over conserved
events.  In the case of activation, both distributions fall exactly on
top of each other and it is hard to distinguish the X's from the
diamonds.  The solid line corresponds to the elasticity scaling,
$N \sim r^{-1.5}$ as discussed in the text.}
\label{num}
\end{figure}

The size of events is usually related to the number of atoms involved in
the rearrangement of the network, i.e., the number of atoms that are
displaced more than a threshold distance $r_c$. In Fig. \ref{num} we
plot the number of atoms involved as a function of $r_c$, for the
conserved and the full set of events.  A local topological rearrangement
will generally push the surrounding atoms outwards, or occasionally pull
them slightly inwards. The distance $\Delta r$ over which the
surrounding atoms are pushed away (pulled inwards) will, because of
elasticity arguments, scale with the distance $r$ from the rearrangement
as $\Delta r \sim r^{-2}$, for sufficiently large $r$. Alternatively,
this can be rewritten as $\Delta r \cdot r^2 = V_e$, where $V_e$ is a
constant volume independent of the distance $r$. The distance over which
the atomic displacement $\Delta r$ exceeds a threshold displacement
$r_c$ is then equal to $r=\sqrt{V_e/r_c}$, and the number of atoms
displaced more than $r_c$ will then scale with $r_c$ as $N_e \sim
V_e^{3/2} r_c^{-3/2}$.  Figure \ref{num} shows that this scaling holds
in the region where $0.1<r_c<1.0$ \AA\, and $N_e \ll N$.  Selecting a
threshold at 0.1 \AA\, the lower bound but also the typical vibration
displacement at room temperature in Si, we find that, on average, about
50 atoms are involved in an event, clearly beyond the very local
mechanism but well below the highest numbers proposed.

\begin{figure}
\epsfxsize=8cm
\epsfbox{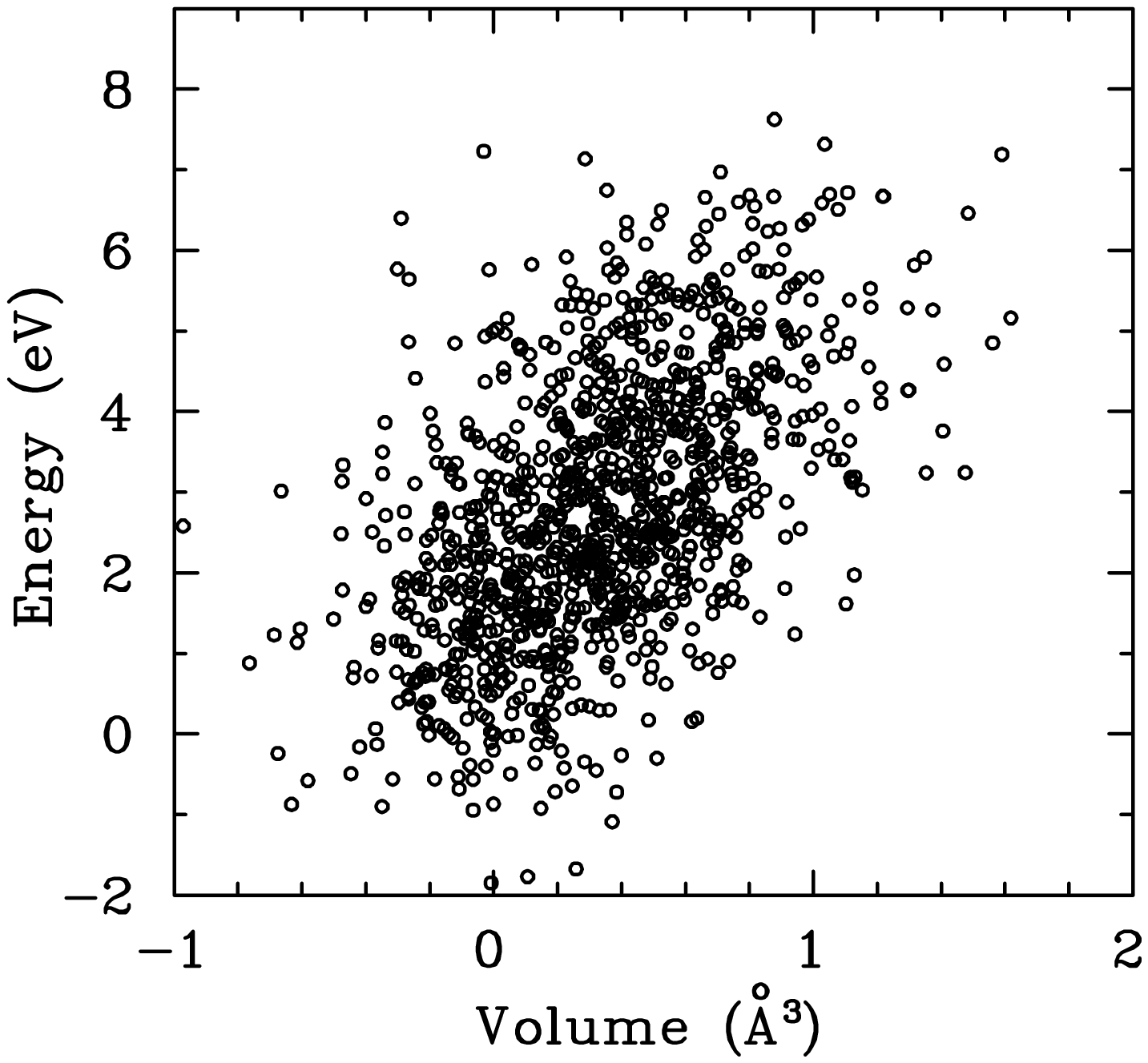}
\epsfxsize=8cm
\epsfbox{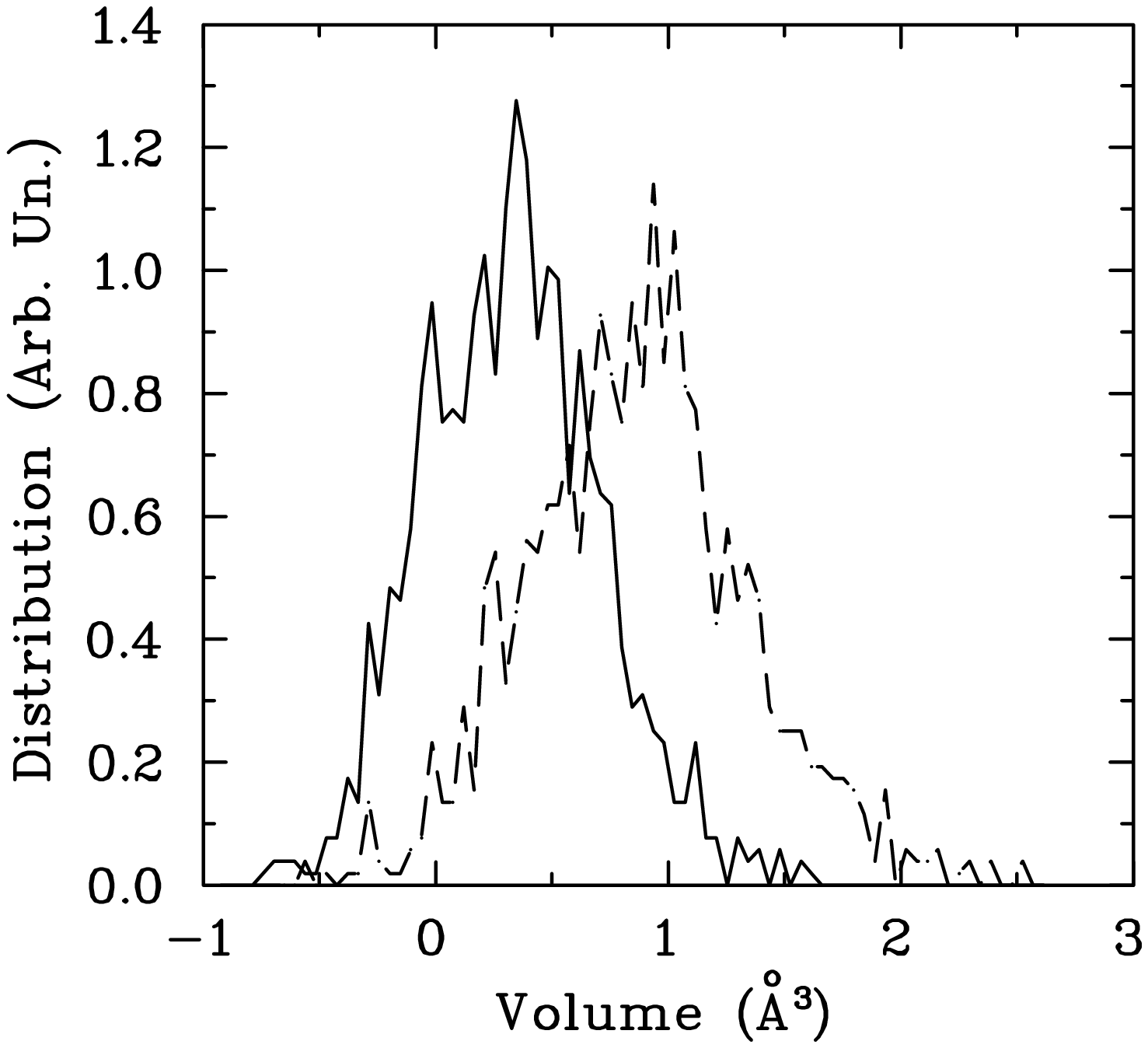}
\caption{Top: Asymmetry energy as a function of the volume of the event
(as defined in the text) for conserved events.  Bottom: Distribution of
the volume of events for conserved events at the saddle point (dotted
line) and at the new minimum (solid line).}
\label{volE}
\end{figure}

The size of the events can thus be measured also without the introduction
of a threshold distance, by measuring $V_e=\Delta r \cdot r^2$, averaged
over some range of $r$. Figure \ref{volE} shows the correlation between
the event volume and the energy barrier and asymmetry. The figure shows
that most proposed events tend to expand the sample locally by about
$V_e \sim 1$ \AA$^3$.  Moreover, although scattered considerable, the
data suggest a linear relation between the energy and the volume
expansion with about 4 eV per \AA$^3$.

\begin{figure}
\epsfxsize=8cm
\epsfbox{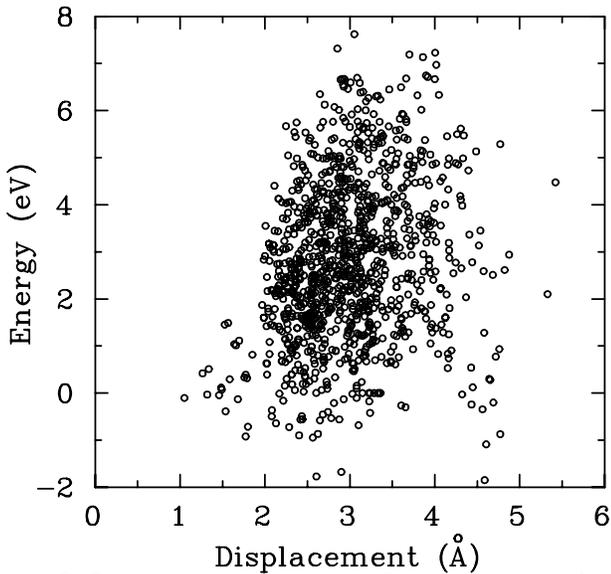}
\caption{Asymmetry energy as a function of total displacement for the
conserved events.}
\label{endisp}
\end{figure}

We can also search for correlations between the total displacement and
the asymmetry energy, which could relate the diffusion length with the
energy. This is plotted, for conserved events, in Fig. \ref{endisp};
there is little correlation. A similar negative result is obtained if
we look at the correlation between the displacement and the activation
energy.

The picture that emerges from these three approaches is that events are
relatively localized, involving around 50 atoms, and require some local
expansion to take place, as would be expected. Correlations between the
size of events and the energy are difficult to establish due, in large
part, to the wide spread of local environment typical of disordered
systems.  These results can be used to put bounds on models of diffusion
and relaxation in amorphous silicon.

\subsection{energetics of coordination defects}
\label{defects}

Weak bonds and coordination defects form another recurring theme in the
study of dynamics and relaxation in {\it a}-Si.  In this section, we
discuss their properties in the sub-set of conserved events. 
The number of bonds broken/created at the saddle point is $3.7 \pm 1.4$
and $3.4 \pm 1.4$ with bond lengths of $2.43 \pm 0.09$ and $2.62 \pm
0.12$ \AA.  At the final point, it is $4.3 \pm 1.6$ (equal number of
bonds created and broken for a conserved event) with respective bond
lengths of $2.46 \pm 0.09$ and $2.55 \pm 0.10$ \AA. 

These numbers are very similar to those obtained by concentrating on
perfect events\cite{barkema98}. This reflects one of our main
conclusions: that correlation between different properties of the
events is weak. If the bond length of the resulting states is typically
longer than that of the initial configuration, it is simply because the
final configurations have typically a significantly higher energy.
There is no preference for a single stretched bond; it is the
medium-range strain that matters.

One would expect a relation between relaxation and defect annihilation,
but how strongly linked these two are is not clear {\em a priori}.
Relatively little is know directly \cite{volkert93}.
To a first approximation, the total energy should follow the density of
defects \cite{wagner96}.  Because of the similarity between crystalline
and amorphous Si, Roorda {\it et al} concluded that relaxation occurs
through defect annihilation~\cite{roorda91}.  Polman {\it et al} find
that Cu diffusion in annealed {\it a}--Si is 2 to 5 times faster than
in non-annealed samples; this increase in diffusion might indicate a
decrease in defects, trapping the Cu \cite{polman90}.
                                                              
\begin{figure}
\epsfxsize=8cm
\epsfbox{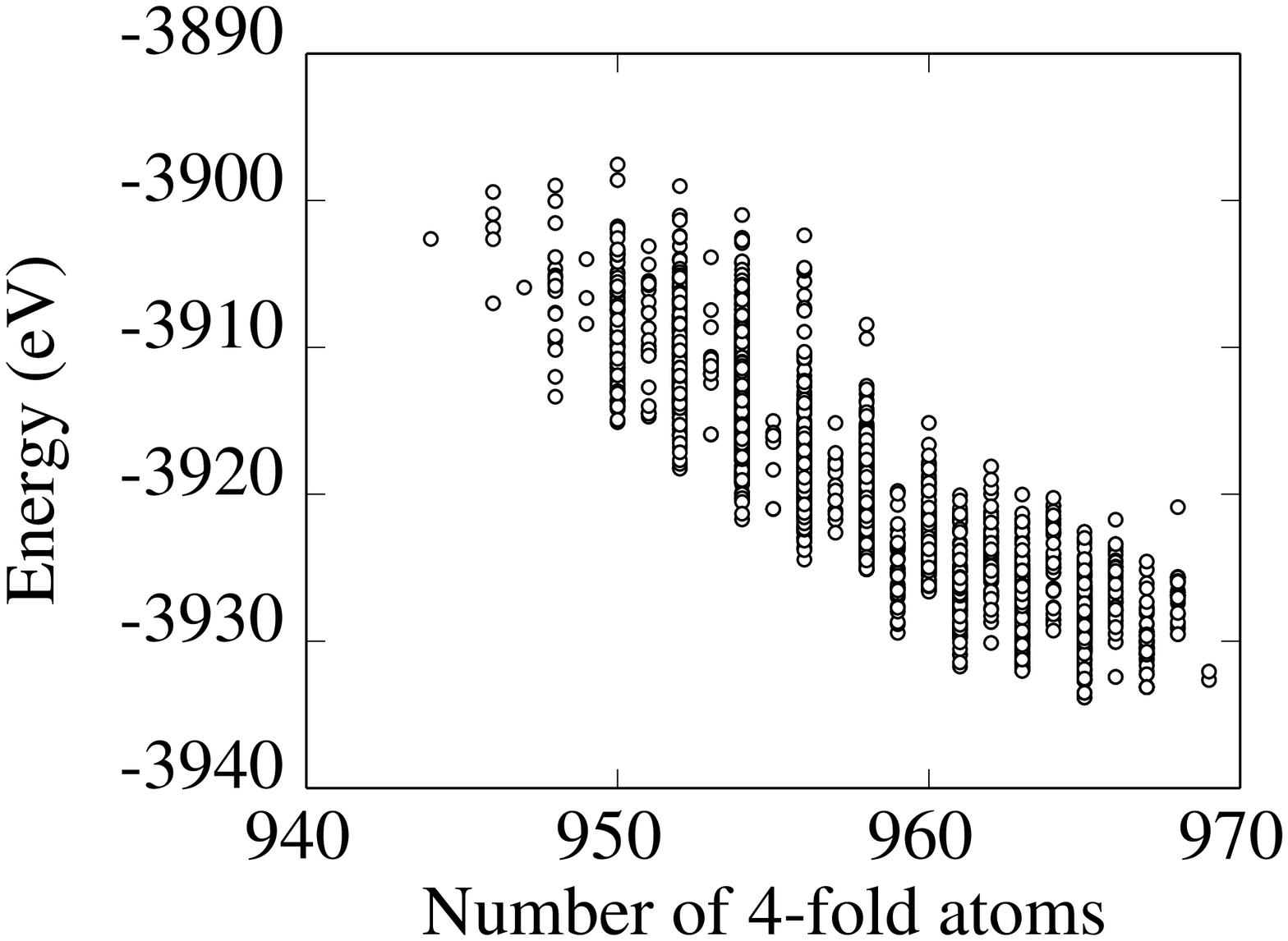}
\epsfxsize=8cm
\epsfbox{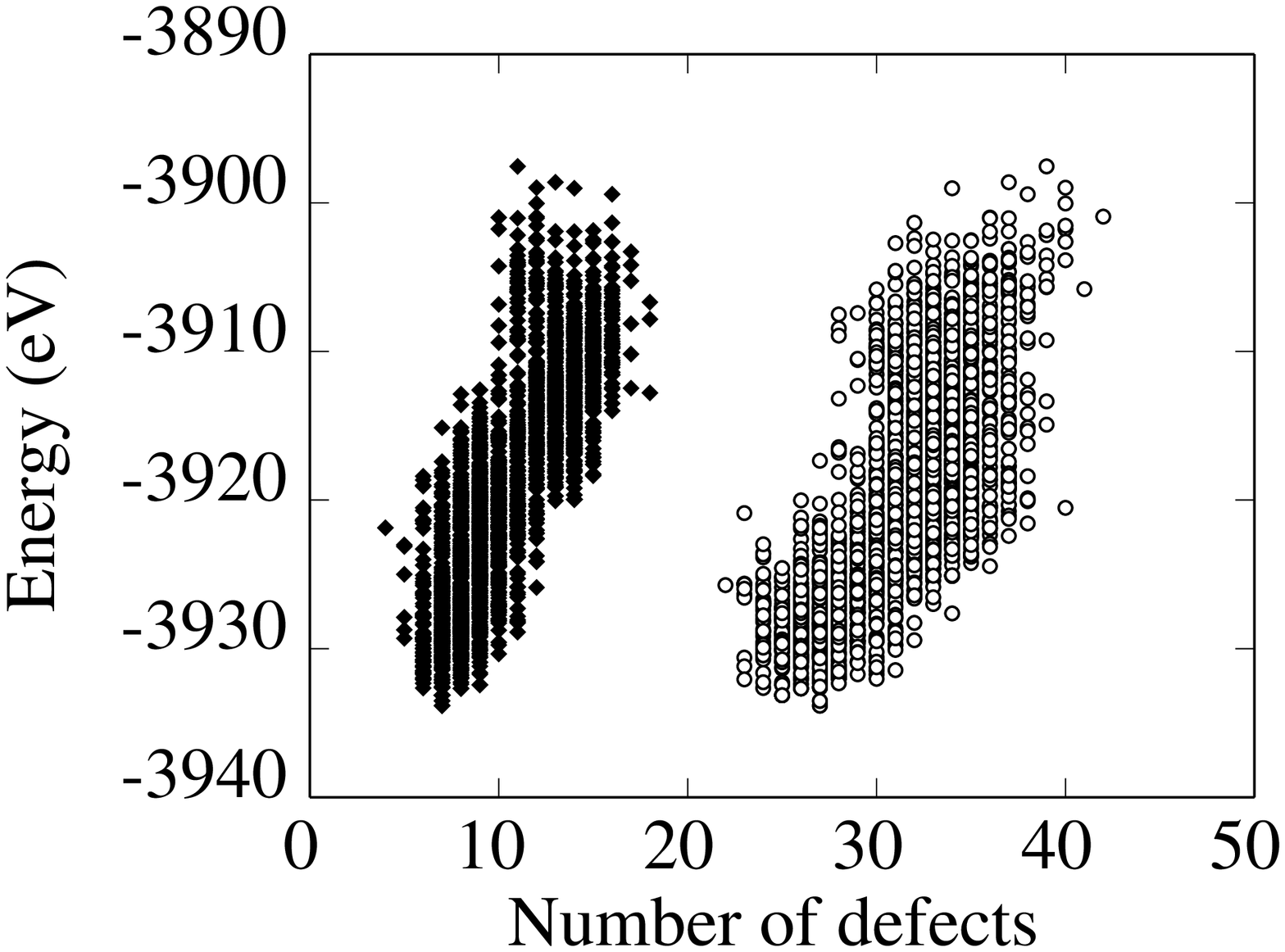}
\caption{Top: Total energy as a function of the number of 4-fold atoms;
bottom: total energy as a function of the number of 3-fold (open circle)
and 5-fold (full diamonds) for all events.}
\label{enedef}
\end{figure}

We see a certain correlation between defects and total energy as
plotted in Figure \ref{enedef}(a).  The distribution of energy for a
given number of defects is wide, going from 12 eV for 35 defects to
about 25 eV for 50 defects.  Looking at the correlations of the energy
with specific type of defects, we find much smaller impacts, with
significant distribution in the total energy for a given number of
3-fold or 5-fold defects as is shown in the bottom panel for Figure
\ref{enedef}.

Clearly, therefore, the definition of defects must also include strained
environment and not just the coordination defects mentioned here ---
relaxing some highly strained ring might require the creation of a bond
defect.  Once again, the situation is much less clear than is generally
thought. We must emphasize here that for less relaxed samples, the
correlation between energy and the number of coordination defects is
much better.

\subsection{Topological classification}
\label{topology}

To go beyond the scalar picture given above, we need to consider in more
detail the nature of the topological changes. The classification scheme
applied here is an extension to the classification scheme that we used
for the perfect events in an earlier letter \cite{barkema98}.  All atoms
that change their neighbors are alphabetically labeled.  the topological
change is determined by specifying the list of all bonds before the
event, and of all bonds after.

For perfect events, there are as many bonds before as after the event,
and moreover, the set of all these bonds can always be organized into a
ring of alternating created and destroyed bonds.  This ring
can be represented by the sequence of atoms visited. For instance, in
event {\it abacbd}, the bonds before and after the event are {\it ab},
{\it ac}, {\it bd}, resp.  {\it ba}, {\it cb}, {\it da}. Thus,
bonds {\it ac} and {\it bd} are replaced by {\it ad} and {\it bc},
while bond {\it ab} is present both before and after the event. Many
equivalent rings exist, but the convention of always using the
alphabetically lowest label makes this classification unique.

If the event is not perfect, this classification scheme has to be
modified.  In case a single bond (or dangling bond) jumps from one atom
to another, the set of bonds does not form a closed ring, but an open
chain of alternating bonds before and after the event; the same
classification scheme can still be used, with the note that the first
and last atom are not bonded.  It also happens that the event comprises
a series of bond exchanges which are in disconnected regions (typically
still nearby - interacting via the strain field). For such events, we
introduce the concept of ``ghost bonds'' (represented by a dot in the
label), that are added to either the set of bonds before or after the
event.

Using this topological analysis, we can have a first crack at the
events. 1148 events in our database have an activation barrier of less
than 8 eV, the other events might be considered unphysical.  Of these,
447 are too complicated (involving too many defects or too many
disconnected regions) to be analyzed, leaving 701 labeled events.

For conserved events, there are no dominant labels, contrary to what is
found for perfect events where three labels account for 85\% of the
events.  Such diversity underscores the difficulty in trying to
identify mechanisms and relating them to experimental information.
Clearly, the dynamics of defects in amorphous silicon is much more
complicated than is usually thought.

In the large set of labels, an often occurring theme is a ring of bonds
corresponding to the Wooten-Winer-Weaire (WWW) bond-exchange mechanism
\cite{wooten85,wooten87}, which in our classification scheme has the label {\it
abacbd}.  We find that up to two WWW moves can take place in a single
event.  This rearrangement changes
the local ring structure and redistributes the strain, affecting the
jump barrier seen by the dangling bond.

After removal of all these rings from the events as such, what
remains is often a single coordination defect that jumps to a 1st, 2nd or
higher neighbor.  Table \ref{tab:eventlabels} presents a partial list of
such labels.  It is remarkable that the diffusion of coordination
defects requires, in general, a topological rearrangement more complex
than one would expect from the displacement of the bonds.  Only nine
events are of the {\it abc} type, the smallest rearrangement possible
for the motion of a bond.

\begin{figure}

\vspace*{-2cm}
\hspace*{-1cm}
\epsfxsize=12cm
\epsfbox{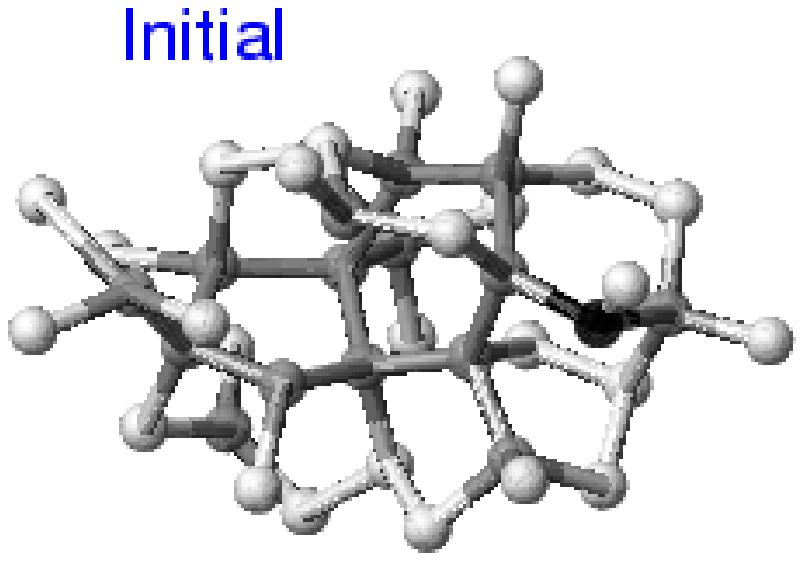}

\vspace*{-2cm}
\hspace*{-1cm}
\epsfxsize=12cm
\epsfbox{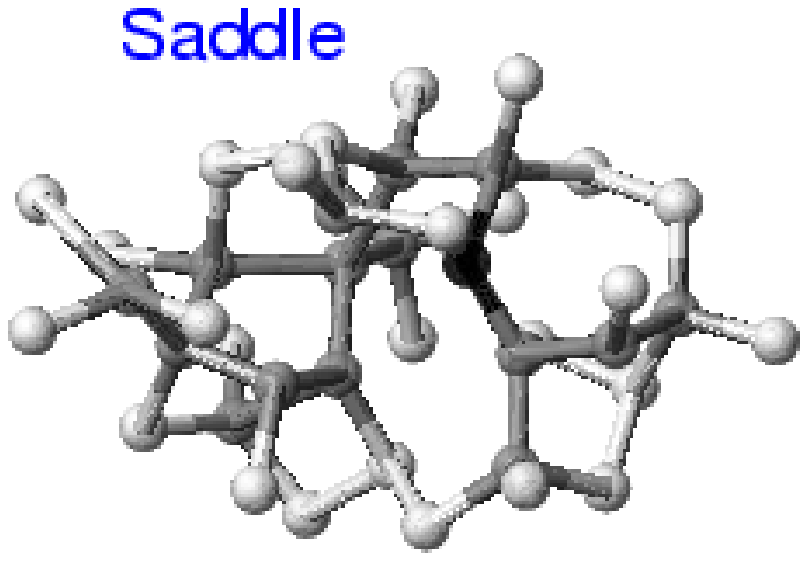}

\vspace*{-2cm}
\hspace*{-1cm}
\epsfxsize=12cm
\epsfbox{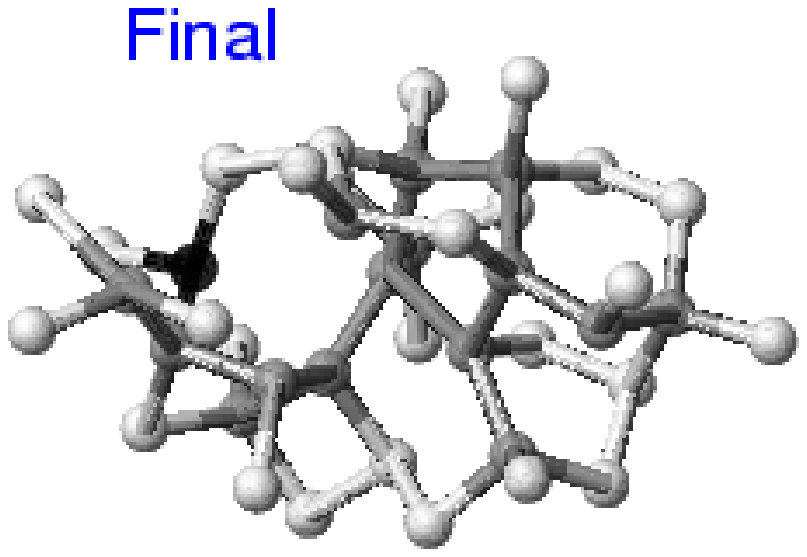}

\vspace*{-1cm}                  
\caption{ Event of type ``abcbdce''. Barrier:
2.8 eV, Asymmetry:  2.1 eV, total displacement 2.5 \AA,
four bonds broken and four created.
}
\label{fig:abcbdce}
\end{figure}

Other events involve longer jumps, at least in topological terms. The
{\it abacbde}, for example, reflect this type of behavior. We show one
such event in  Fig. \ref{fig:abcbdce}. Very few of the classified
events displace more than one defect. (It could be that in the
non-classified events this situation occurs more often.)

\section{Conclusions}

Based on an extensive list of events in {\it a}-Si, representing a wide
range of characters, we can provide a general overview of the nature of
the microscopic mechanisms responsible for the diffusion of topological
defects in this material. To do so, we havei \hfill concentrated on a class of

\begin{table}
\caption{Most common events for low energy conserved events. 
The first column gives the classification label in terms described
above. The second and the third columns refer to the number of events
carrying this label and the number of topological defects --- typically
dangling bonds --- displaced in the event, respectively. The last column,
finally, gives a topological indication of how far the defect
diffuses. A local jump brings a defect from one atom to its
near-neighbor; non-local events will involve jumps to the second-,
third-, or even fourth-neighbor.  \label{tab:eventlabels}}
\begin{tabular}{l|c|c|l}
topology & occurence & no. defects & local/non-local \\ \hline
abacadefbfgbedc & 2 & 1 & local \\
abacadefegfbedc & 2 & 1 & local \\
abacb           & 9 & 1 & local \\
\\
abacbdaeb       & 5 & 1 & local \\
abacbdaebfagb   & 6 & 1 & local \\
abacbdaebfagc   & 2 & 1 & local \\
abacbdaebfbgf   & 4 & 1 & 2nd neighbor \\
abacbdaec       & 8 & 1 & local \\
abacbdaef       & 6 & 1 & local \\
abacbdaefageg   & 2 & 1 & local \\
\\
abacbdbed       & 5 & 1 & 2nd neighbor \\
abacbdbedcfcfg  & 6 & 1 & local \\
abacbdbedcfcgdf & 2 & 1 & local  \\
abacbdbefegfdg  & 4 & 1 & local \\
abacbde         & 3 & 1 & 2nd neighbor \\
abacdcedb       & 8 & 1 & local \\
abacbdedfeb     & 5 & 1 & local \\
abacdcedbfdfe   & 2 & 1 &2rd neighbor \\
abacdcedbfdfg   & 8 & 1 &3rd neighbor \\
abacdcedbfdfgeg & 4 & 1 &3rd neighbor \\
abacdcedfcb     & 4 & 1 & local \\
abacdcefbfgbechdi & 3 & 1 & 3rd neighbor \\
abacdebefbdcg   & 11 & 1 & 2nd neighbor \\
abacdefegfdcb   & 3 & 1 & local \\
abc             & 9 & 1 & local \\
abcde           & 2 & 1 & 4th neighbor \\
abcbdcebf       & 7 & 1 & local \\
abcbdcebfcg     & 5 & 1 & 3rd neighbor \\
abcdc.ec.fdg.ehi & 1 & 4 & 2nd neighbor 
\end{tabular}
\end{table}

\noindent
events that involve the displacement of coordination defects while
keeping their overall number constant.

Analyzing a wide range of structural and topological properties of these
events we find that: (1) In a well-relaxed sample, there is little
correlation between the number of defects and the total energy; the
relaxation of strain can take place in more subtle ways, sometimes
involving the creation of topological defects.  (2) Taken into account
the use of an empirical potential,  the activation barriers are in
agreement with experimental value. (3) We find little correlation
between the activation barrier or the asymmetry and the deformation of
the network, either in terms of the number of atoms involved or the
total displacement; the energy is best described in term of the volume
of an event. (4) A topological analysis of the conserved events show an
unexpected richness; we find literally hundreds of different mechanisms
that cannot easily be put in a few classes. As a rule, the
Wooten-Weaire-Winer bond exchange mechanism, dominant for perfect
events, still plays a major role. Defect diffusion is often local, with
coordination defects jumping from one atom to a neighbor, but it can also
go as far the fourth neighbor, in a chain-like fashion. 

These results can be used to put bounds on models of diffusion and
relaxation in amorphous silicon. For example, the Fedders and Branz
model for {\it a}--Si:H \cite{fedders95} states that (1) relaxing the
defect structure often requires several atoms to move simultaneously,
(2) only by cooperative motion do the position changes lower or conserve
the total energy, (3) the size of the barrier generally increases with
the number of atoms that must move simultaneously. Our results support
points (1) and (2) but not (3).  

This study represents only a first step in the study of microscopic
activated mechanisms in {\it a}-Si. More work remains to be done to
converge barriers using more accurate interaction potentials. It is
important also to try to connect some of these results with hard
experimental numbers, a challenge both for theorists and
experimentalists. Already, however, we can see that the dynamics of
disordered materials promises to be much more complicated than was
thought before.

\section*{Acknowledgements}

N.M. acknowledges partial support from the NSF under grant number
DMR-9805848 as well as generous time allocations of the computers of
the High Performance Computing Center at Delft Technical University,
where part of the analysis was done. GB acknowledges the High
Performance Computing group at Utrecht University for computer time.

\vspace*{-0.3cm}

\bibliographystyle{prsty}

\begin{thebibliography}{99}

\vspace*{-1.0cm}

\bibitem[(a)]{mousadd} email: mousseau@helios.phy.ohiou.edu.

\bibitem[(b)]{barkadd} E-mail: barkema@phys.uu.nl.

\bibitem{sputt} X. Liu, B. E. White Jr., R. O. Pohl, E. Iwanizcko, K.
M. Jones, A. H. Mahan, B. N. Nelson, R. S.  Crandall, and S. Veprek,
Phys. Rev. Lett. {\bf 78}, 4418 (1997).

\bibitem{roorda91} S. Roorda, W. C. Sinke, J. M. Poate,
D. C. Jacobson, S. Dierker, B. S. Dennis, D. J. Eaglesham, F. Spaepen, 
and P. Fuoss, Phys. Rev. B {\bf 44}, 3702 (1991).

\bibitem{mousseau97} N. Mousseau and L. J. Lewis, Phys. Rev. Lett. {\bf
78}, 1484 (1997).

\bibitem{barkema96}
G.T. Barkema and N. Mousseau, Phys. Rev. Lett. {\bf 77}, 4358 (1996).

\bibitem{weber85} T. A. Weber and F. H. Stillinger, Phys. Rev. B {\bf 32},
5402 (1985).

\bibitem{barkema98} G.T. Barkema and N. Mousseau,
Phys. Rev. Lett. {\bf 81}, 1865 (1998).

\bibitem{numrec} W.H. Press et al., {\it Numerical Recipes},
Cambridge University Press, Cambridge, 1988.

\bibitem{assumption} We have recently relaxed this assumption by
introducing some trailing in the algorithm. Such modification has been
significant in metallic glasses and polymers but has not changed the
results for {\it a}-Si.

\bibitem{doye97} J. P. K. Doye and D. J. Wales, Z. Phys. D {\bf 40}
466 (1997).

\bibitem{mousseau98}
N. Mousseau and G.T. Barkema, Phys. Rev. E {\bf 57}, 2419 (1998);
Comput. Sci. Eng. {\bf 1}, 74-80 (1998).

\bibitem{ding86}  K. Ding and H. C. Andersen, Phys. Rev. B {\bf 34}, 6987
(1986)

\bibitem{hauch99} J. A. Hauch, D. Holland, M. P. Marder and H. L.
Swinney, Phys. Rev. Lett. {\bf 82}, 3823 (1999).

\bibitem{mousseau97b} N. Mousseau and L. J. Lewis, Phys. Rev. B {\bf
56}, 9461 (1997).

\bibitem{rdf-exp}
G. Etherington, A. C. Wright, J. T. Wenzel, J. C. Dore, J. H. Clarke and
R. N. Sinclai, J. Non-Cryst. Sol. {\bf 48}, 265 (1982).

\bibitem{bazant} J. F. Justo, M. Z. Bazant, E.  Kaxiras,V. V. Bulatov,
S. Yip, Phys. Rev. B {\bf 58}, 2539 (1998).

\bibitem{custer94} J. S. Custer, M. O. Thompson, D. C. Jacobson,
J. M. Poate, S. Roorda, W. C. Sinke, and F. Spaepen,
Appl. Phys. Lett. {\bf 64}, 437 (1994).

\bibitem{williamson95} D. L. Williamson, S. Roorda, M. Chicoine,
R. Tabti, P. A. Stolk, S. Acco and F. W. Saris,
Appl. Phys. Lett. {\bf 67}, 226 (1995).

\bibitem{laaziri}
K. Laaziri, S. Kycia, S. Roorda, M. Chicoine, J. L. Robertson,
J. Wang, and S. C. Moss , Phys. Rev. Lett. {\bf 82},  3460 (1999).

\bibitem{simulations} For a set of references, see
L.J. Lewis and N. Mousseau, Comput. Mat. Science {\bf 12}, 210-241
(1998).

\bibitem{shin93} J. H. Shin and H. A. Atwater,
Phys. Rev. B {\bf 48}, 5964 (1993).

\bibitem{muller94} G. M{\"u}ller, G. Kr{\"o}tz, S. Kalbitzer and
G. N. Greaves, Phil. Mag. B {\bf 69}, 177 (1994).

\bibitem{masson95} D. P. Masson, A. Ouhlal and A. Yelon,
J. Non-Cryst. Sol. {\bf 190}, 151 (1995).

\bibitem{liang94} Z. N. Liang, L. Niesen, G. N. van den Hoven, and
J. S. Custer, Phys. Rev. B {\bf 49}, 16331 (1994).

\bibitem{polman90} A. Polman, D. C. Jacobson, S. Coffa, and J. M. 
Poate, Appl. Phys. Lett. {\bf 57}, 1230 (1990).

\bibitem{volkert93} C. A. Volkert,
J. Appl. Phys. {\bf 74}, 7107 (1993).

\bibitem{wagner96} S. Wagner, H. Gleskova, and J-I. Nakata,
J. Non-Cryst. Sol. {\bf 198-200}, 407 (1996).

\bibitem{wooten85} F. Wooten, K. Winer, and D. Weaire, Phys. Rev. Lett. {\bf
54}, 1392 (1985).

\bibitem{wooten87} F. Wooten and D. Weaire, Solid State Phys. {\bf 40},
1 (1987).

\bibitem{fedders95} P. A. Fedders and H. M. Branz,
J. Non-Cryst. Sol. {\bf 190}, 142 (1995). 

\end{thebibliography}

\end{document}